\documentstyle[12pt]{article}
\begin{document}

\vskip 5cm
{\bf{\LARGE Two-Dimensional Seven-State Potts Model \\ \\
 Under External Magnetic
Field}}
\vskip 1cm
\begin{tabbing}
xxxx\=xxxxxxxxxxxxxxxxxxxxxxxxxxxxxxxxxxxxxxxxxxxxxxxxxxxxxxxxxxxxxxxxxxxxxxxxxxxxxxxxxxx\kill
       \>{\bf T. \c{C}EL\.{I}K, Y. G\"{U}ND\"{U}\c{C} and M. AYDIN} \\
       \>   \\
       \> {\it Hacettepe University, Physics Department} \\
       \> {\it 06532 Beytepe, Ankara, Turkey }\\
\end{tabbing}
{\small
{\vskip 1cm}
\begin{tabbing}
xxxx\=xxxxxxxxxxxxxxxxxxxxxxxxxxxxxxxxxxxxxxxxxxxxxxxxxxxxxxxxxxxxxxxxxxxxxxxxxxxxxxxxxxxxxxxxxxxx\kill
       \>{\bf Abstract.-} The two-dimensional Potts Model with seven states
under external field\\
       \>is studied using a cluster algorithm. Cluster size distribution and
the fluctuations in\\
       \>the average cluster size  provide helpful information on the order of
phase transitions.\\
\end{tabbing}
}

Potts model \cite{Potts:1952} is known to have a very rich critical behaviour
and considered as a testing ground for both analytical and numerical methods.
Recently, the problem is mainly focused on determining the nature of the phase
transition, especially the weak first-order transition occuring in this
model \cite{Billoire:1995, Lee:1990}. In two dimensions, the Potts model
displays a second-order phase
transition for the number of states $q = 2, 3$ and $4$, and first-order
transition for $q \geq 5$, where transitions become stronger as $q$
increases \cite{Baxter:1973, Wu:1982}.  One of the most reliable methods to
study
first-order transitions is to observe the doubly-peaked probability
distribution
for energy \cite{Binder:1984, Binder:1986}. At the critical point,
the order parameter and the cluster size distribution should also
exhibit the same behaviour.
A  first-order phase transition however is expected to become weaker
under an increasing external field and it becomes a difficult
task to recognize a double-peak in a weak first-order phase transition.
Further increase in external field reduce the strength of the transition and
the phase transition disappears for large external field strengths.
This process is the result of two competing interactions: spins form sizable
clusters in some regions because of the tendency to align in the field
direction, while the thermal fluctuations result in disintegration of the
clusters.  The external field yields the ordering of the system at
higher temperatures, shifting the critical temperature to higher values
as well as reducing fluctuations in the system.
On the other hand, the dynamical changes associated with a phase
transition in the system are reflected in the fluctuations and variations
in the cluster size.
The aim in this work is to observe, by using a cluster algorithm, the dynamical
behaviour of cluster size variations with respect to temperature and the
external
field.

$\;$

In the present work, the
critical behaviour of the two-dimensional Potts model with seven states
under external magnetic field is studied using cluster algorithm.
Equilibrium averages, fluctuations in the average cluster size and the
histograms for energy, order parameter and the average cluster size are
obtained as a function of the temperature for different field strengths.

$\;$

The Hamiltonian of the two-dimensional Potts model is given by
\begin{equation}
     {\cal H} = K \sum_{<i,j>} \delta_{\sigma_{i},\sigma_{j}} + H \sum_i
\delta_{\sigma_{i},o}.
\end{equation}
Here $K=J/kT$ ; where $k$ and $T$ are the
Boltzmann constant and the temperature respectively, and $J$ is the magnetic
interaction between spins $\sigma_{i}$ and $\sigma_{j}$, which can take
values $0,1,2, ..., q-1$ for the $q$-state Potts model and $H=h/kT$ with $h$ is
the external field along the orientation 0.
Reader can refer to the review article by Wu \cite{Wu:1982} for detailed
information about the model.
Order of the transitions can be studied by calculating specific heat

\begin{equation}
C=\frac {1}{kT^{2}} (<E^{2}>-<E>^{2})
\end{equation}
and the Binder cumulant \cite{Binder:1981}
\begin{equation}
B=1-\frac {<E^{4}>}{3<E^{2}>}
\end{equation}
on finite lattices, where $E$ is the energy of the system.

$\;$

The dynamical evolution of clusters and fluctuations in the observables
are strongly dependent on the correlations in the system. When the
temperature is high the clusters are small, and they start growing as the
critical point is approached.
If the correlation length in the system is finite (especially when it is
shorter than the lattice size), existing large clusters may break down to
smaller ones with thermal fluctuations in the system. Hence in such a
system, very large and very small clusters may coexist, resulting in large
fluctuations in the cluster size. If the correlation length is very large
as in the case of weak first-order phase transition or infinite as in
the case of a second-order transition, the existing large clusters
can not easily disintegrate with thermal fluctuations. A similar effect
can be seen when an external field is turned on in a system with first-order
phase transition. The effect of the external magnetic field aligns
the spins parallel to the field direction creating large clusters and
preserve these clusters from the effects of thermal fluctuations. These
considerations led us to the usage of the average cluster size,
fluctuations in the cluster size, cluster size distribution as the
operators to investigate the changes in the phase transition with respect
to variations of temperature and external
field strength. The cluster algorithm
which introduces a global update by means
of selecting clusters and updating all spins in the cluster proved to be
very successful in eliminating critical slowing down and super critical
slowing down as in the first-order
transitions in systems such as Potts \cite{Swendsen:1987} and
O(N) \cite{Wolf:1989} models. Studying any
operator which uses information related to the clusters is extremely
straightforward since this information is inherited in the cluster
algorithm.  The algorithm used in this work is similar to Wolf's
algorithm, with the exception that, $H$ is incorporated following  Dotsenko et
al
\cite{Dotsenko:1991} and before calculating the observables,
searching the clusters is continued until the total number of sites in all
searched clusters is equal to or exceeds the total number of sites in the
lattice.

$\;$
\begin{figure}[ht]
\vspace{6.0cm}
\includegraphics{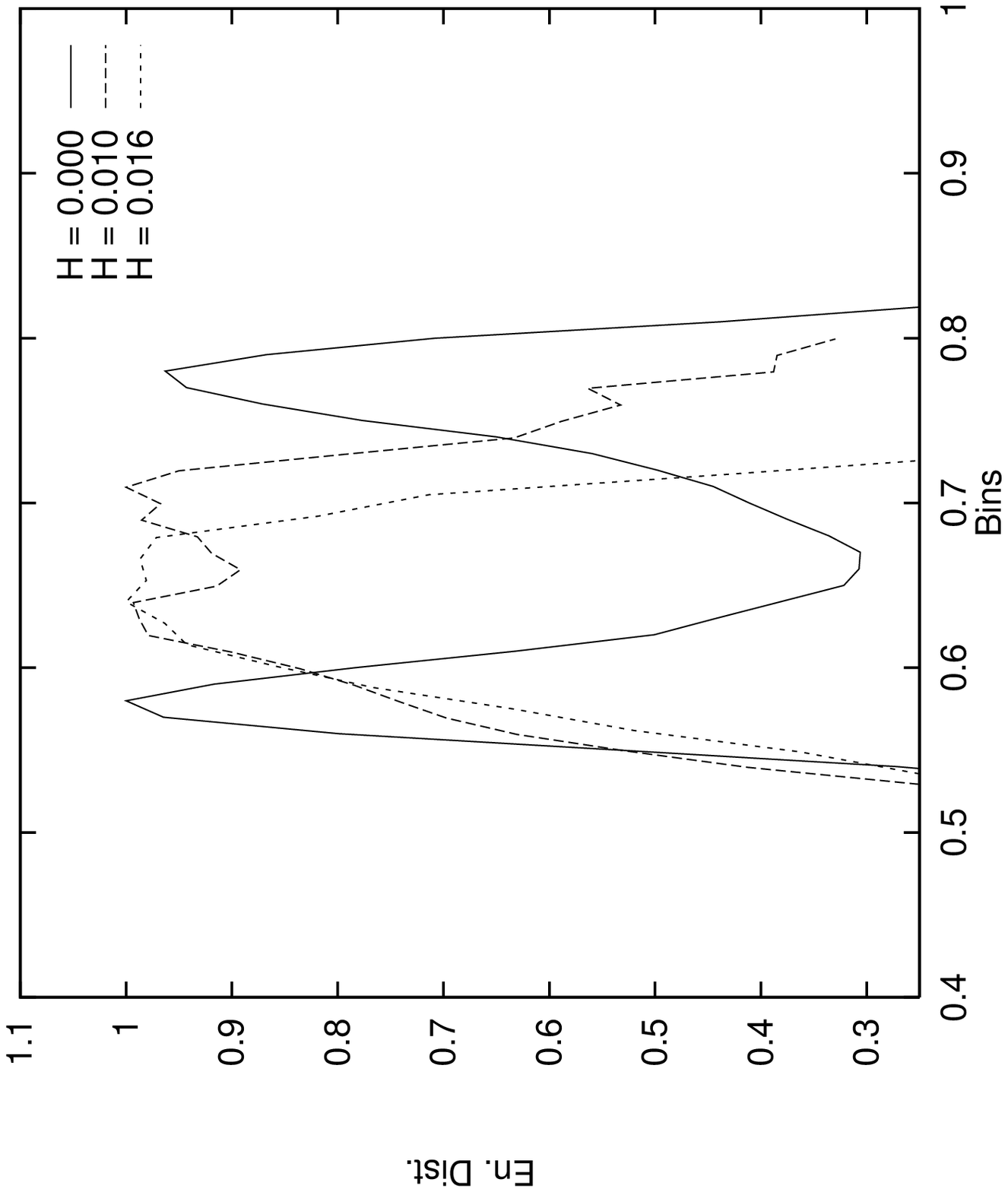}
\includegraphics{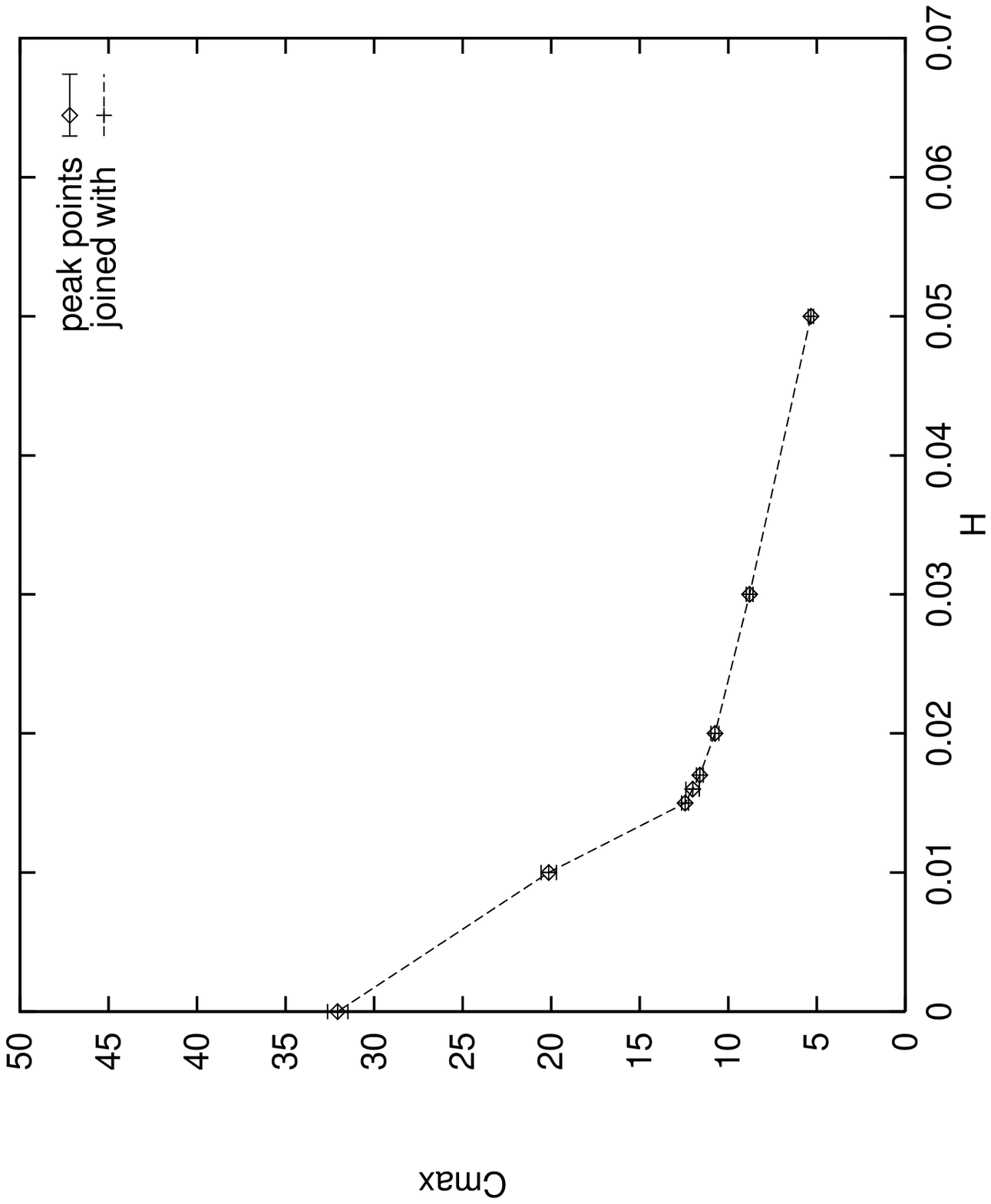}
\begin{tabbing}
xxxxxxxxxxxxxxxxxxxxxxxxxxxxxxxxxxxxxxxxx\=xxxxxxxxxxxxxxxxxxxxxxxxxxxxxxxxxxxxxx\kill
Fig. 1 . {\small Energy histograms at the critical } \>Fig. 2 . {\small
Specific heat peaks vs. $H$.} \\
{\small point $K_c$ for several values of $H$.}
\>{\small Dashline is drawn to guide the eye.} \\
\end{tabbing}
\end{figure}
In this work, the two-dimensional, $q = 7$ Potts model with external
magnetic field has been simulated on $64 \times 64$ square lattice. After
thermalization with $10^{5} - 2 \times 10^{5}$ sweeps, $5 \times 10^{5} -
2 \times 10^6$ iterations are performed at different values of the
coupling $K$ and the field $H$.  Longer runs with up to $3 \times 10^{6}$
iterations are done near the finite-size critical value $K_{c}$ at each
field strength $H$. For $H = 0.00, 0.01, 0.015, 0.016, 0.02, 0.03$
 and $0.05$ the long
runs are performed at $K_{c} = 1.2909, 1.2774, 1.2707, 1.2696, 1.2652, 1.2531$
and $1.2299$ respectively, where these points
are chosen as the estimated peak positions of the specific heat. From
these long runs we have evaluated by appropriate reweighting
\cite{Ferrenberg:1988}
the energy as a function of the temperature for each value of $H$.
The continuous curves of energy, specific heat and the
Binder cumulant obtained through extrapolation from one long run are in
very good agreement with the data points over rather wide ranges of $K$
for each $H$ value . The point should be stressed here is that the success
of an extrapolation depends on number of statistically
independent configurations used, which is provided by the cluster
algorithm employed here.
As a first criterion to distinguish phase transition, we have
checked  for each value of $H$, the energy distribution at the critical
temperature where the specific heat possess a peak. At $H = 0$,
where seven-state Potts model exhibits first-order character,
a distinct double-peak is observed. As one can see from figure 1,
with increasing $H$ the double peak
converges to a single peak while $H$ reaches to the value of 0.016.
Our simulations indicate that $H = 0.016$ is almost at the verge of the
first-order phase transition region and for larger values of $H$
 we have seen no double
peak behaviour but from the observation of a single gaussian energy
distribution, it is hard to conclude whether the transition becomes
second-order or it is totaly wiped out.

$\;$
\begin{figure}[htb]
\vspace{6.0cm}
\includegraphics{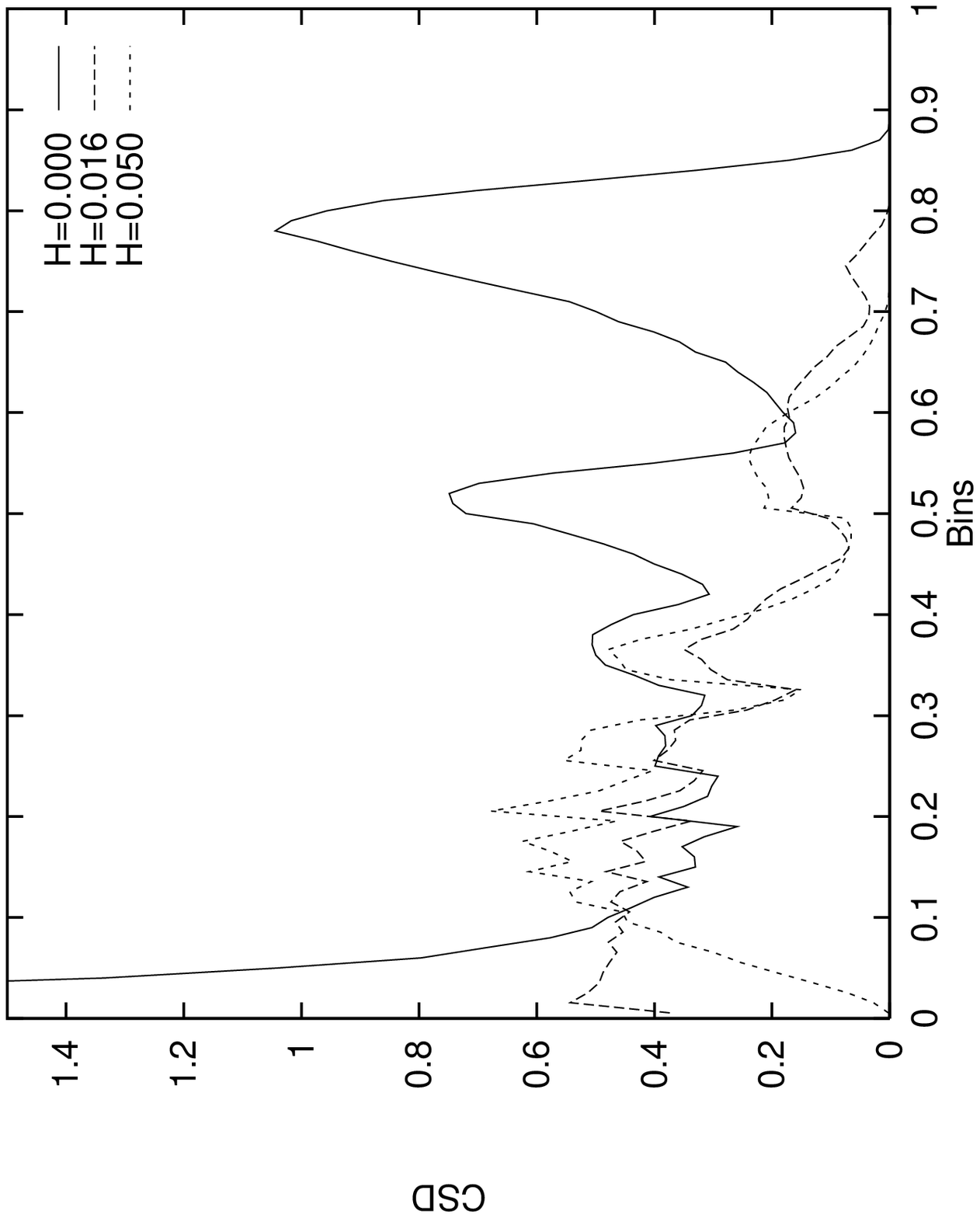}
\includegraphics{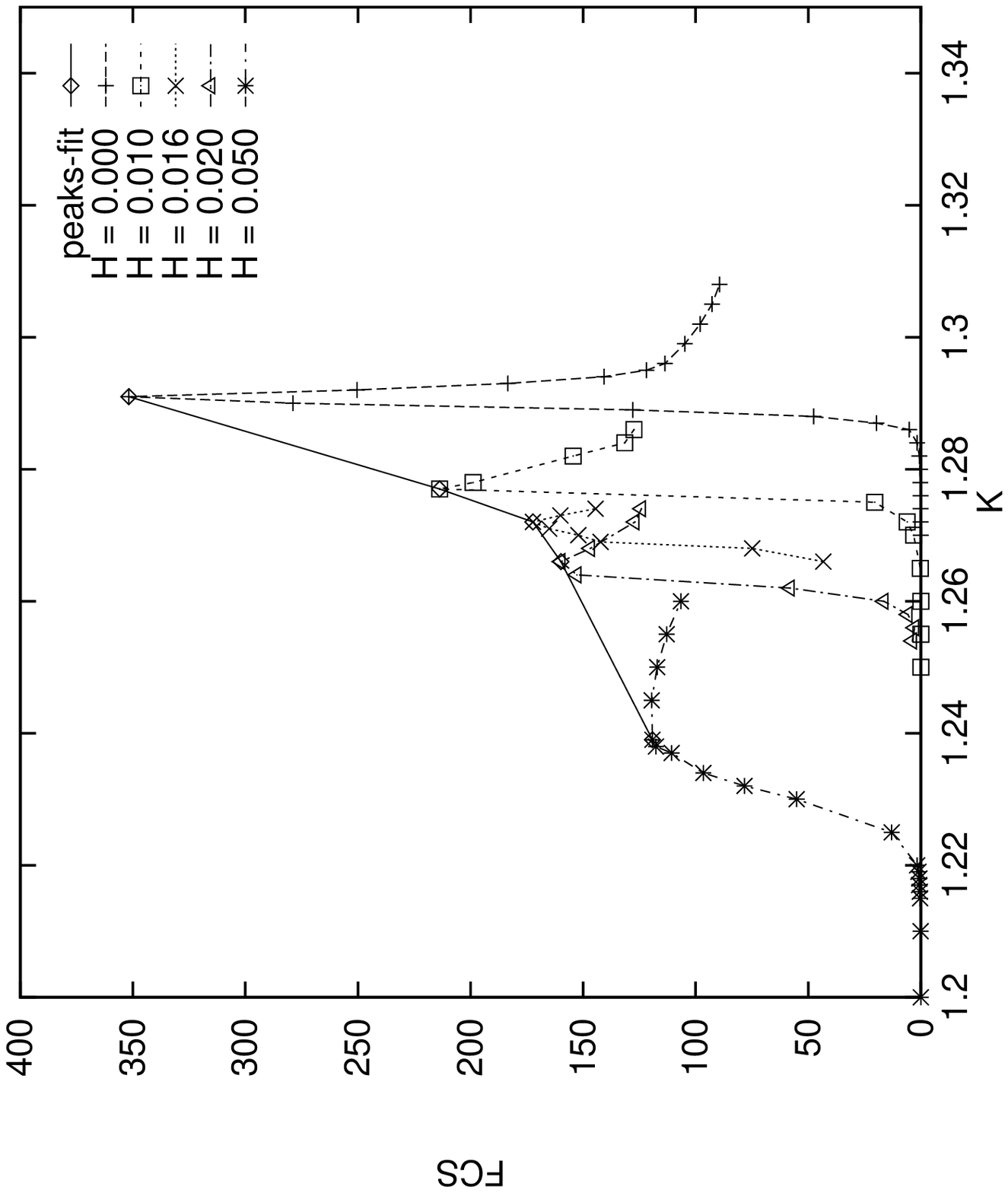}
\begin{tabbing}
xxxxxxxxxxxxxxxxxxxxxxxxxxxxxxxxxxxxxxxxx\=xxxxxxxxxxxxxxxxxxxxxxxxxxxxxxxxxxxxxx\kill
Fig. 3 . {\small Cluster size distributions for seve-}     \>Fig. 4 . {\small
Fluctuations in cluster size} \\
{\small ral values of $H$.}                              \>{\small vs. $K$.} \\
\end{tabbing}
\end{figure}
{}From specific heat peaks and the Binder cumulant minima, relevant
information for the phase transition may also be obtained. For a first-order
phase transition, the specific heat exhibits Dirac-delta function like
shape. While the system moves towards the region of softer phase
transitions with increasing $H$, the peak widens as well as the height is
reduced. In the case of no phase transition, the system may still be expected
to possess some fluctuations in the energy.
For the set of values of the increasing external field, one obtaines a series
of
specific heat curves which become lower and wider as one moves towards the
lower valus of $K$.
Instead of a polynomial fit, here we have simply joined the tips of the
successive specific heat curves by straight lines and displayed it in figure 2.
This most
naive presentation already prevails the existence of
two distinct regimes and the turnover point is at $H=0.016$ where
the line of first order transitions comes to an end.
Our data on the Binder cumulant minima led to the same conclusion.

$\;$

In order to have more microscopic insight and demonstrate what possibly the
cluster search may add to the knowledge, we studied the time evolution
of average cluster size (CS), its fluctuations (FSC)
and the cluster size distributions (CSD).
While the average energy fluctuates within a limited range about a mean value,
the size of then existing clusters may easily fluctuate between the two
extremes, namely from a single spin to a cluster of lattice size.
Hence it seems to us that one may draw more information from cluster
formation about the dynamical changes occuring in the system.
Although the clusters we consider are of the Swendsen-Wang type, since
the changes in cluster size are due to the update algorithm used,
the observables not dealing with the exact cluster size (like
$FCS$ and $CSD$) should give correct information on the system.
We have studied the time evolutions over 2 - 3 million iterations
of average cluster sizes about $K_c$ for each value of the external
field. These  data are very much the same of the corresponding
time-series for energy but more amplified and displays more insight
about the grouping of spins leading to the concerned
energy value. Two-state structures of first-order transitions and the long
range fluctuations of second order phase transitions are more easily
detectable comparing to the energy-time series.
In figure 3, we show the corresponding cluster size distributions in terms
of histograms over the cluster size (normalized to the volume) in the range of
0 to 1.
{}From figure 3, one can easily follow the appearence of the two-state of
small and large size clusters as for the $H=0.00$ case,
almost equal weight for all sizes
as for $H=0.016$ and loosing the small and large  clusters in favor of
medium sizes for $H > 0.016$.  The point to bring into the readers attention
here is that the distribution of small clusters (bins $<$ 0.1) are very
different
for the three cases considered. For a first-order phase transition a
substantial
amount of small clusters are present  and they gradually disappear while the
phase transition weakens.

$\;$

The fluctuations $FCS$ ($FCS=<(CS)^{2}>-<CS>^{2}$) in the average cluster
size  are calculated for different values of $H$ , as a function of
$K$. For the case of no external field, at small $K$
values (high temperatures) the cluster sizes are small, hence $FCS$ is small.
As $K$
increases, the cluster sizes are getting larger, but $FCS$ increases due
to the existence of the small-size clusters as well as the large ones. At
the critical point, formation of the largest clusters leads to the
largest value of $FCS$. One can make a similar discussion for $K > K_{c}$
(when the critical point is approached from above).
When the external field is
turned on, spin alignments in the field direction reduces the fluctuations
and the probability of finding large and small clusters in coexistence
decreases with the increasing magnetic field. $FCS$ for
$H=0.00, 0.01, 0.016, 0.02$ and $0.05$ are plotted in figure 4.
Joining the tips of the successive $FCS$ curves in figure 4 yields nothing but
exactly what is displayed in figure 2
apart from a small shift in $K$ due to the finite size effects.
As can be seen from
these plots, increasing magnetic field reduces the critical temperature,
as well as the peak heights. As $H$ increases, the $FCS$ curves widen and
after some $H$ value, it is hard to see a distinct maximum point.
At $H=0.05$ the $FCS$ curve is step function like where the phase
transition is already wiped out.
Besides the appearence of two  different regimes on either side of the
field value $H=0.016$, figure 4 also displays differences in the distinct
shapes of the $FCS$ curves belonging to two regimes. We observed that the
$FCS$ curves for $H > 0.016$ are continuous and a polynomial fit is
possible which may be denoted as the sign of a smooth change in the average
cluster size in the system.
The curves for $H < 0.016$ have the shape of a spike, for which the only
possible polynomial fit is to have two distinct polynomials
joining at the tip. This appearent discontinuity in the rate of
fluctuations in cluster size may be attributed to the two state structure
of first-order phase transitions.
The low temperature asymptotic values for all $H$ seem to be almost
the same regardless of the order of transition. Because of small thermal
fluctuations at low temperatures, the clusters freeze  after they are
formed, resulting in a small $FCS$ value.

$\;$

In conclusion, one can see from $CSD$ and $FCS$ plots
and from the energy histograms obtained near $K_{c}$ for each considered
value of $H$ and from the extrapolations performed using the fitted values of
the maximums of the specific heat, the first-order transition in 2D $q = 7$
Potts model seems to disappear at $H=0.016$.
What we would like to bring into attention here is that
the observation of the temperature
dependence of fluctuations  in the average cluster size and the
cluster size distributions  gives valuable information  about the
nature of the transition and this kind of investigation may enrich the
physical insight when employed in studying phase transitions.

$\;$

Further work is planned to study random-field and random-bond Potts models
using the same algorithm.
$\;$
\begin{center}
$\star  \star  \star$
\end{center}

The support from T\"{U}B\.{I}TAK through project TBAG-1299 is acknowledged.

\vskip 2cm

\pagebreak

\pagebreak

\centerline{FIGURE CAPTIONS}

\begin{description}
\item {Figure 1.} Energy histograms at the critical point $K_c$ for several
values of $H$.

\item {Figure 2.} Specific heat peaks vs. $H$. Dashline is drawn to guide the
eye.

\item {Figure 3.} Cluster size distributions for several values of $H$.

\item {Figure 4.} Fluctuations in cluster size vs. $K$.
\end{description}
\end{document}